\documentclass{aa}
\usepackage{graphicx}
\usepackage{natbib}
\usepackage{amsmath}
\bibpunct{(}{)}{;}{a}{}{,}

\newcommand{\xb}{\boldsymbol{x}}
\newcommand{\xbo}{\boldsymbol{x_o}}
\newcommand{\bb}{\boldsymbol{b}}
\newcommand{\Ob}{\boldsymbol{0}}
\newcommand{\qb}{\boldsymbol{q}}
\newcommand{\db}{\boldsymbol{d}}
\newcommand{\Dc}{\mathcal{D}}
\newcommand{\shm}{{\it SHM}}

\begin{document}

\title{On Optimal Detection of Point Sources in CMB Maps}

   \author{R. Vio\inst{1}
          \and
          L. Tenorio\inst{2}
          \and
          W. Wamsteker\inst{3}
          }

   \offprints{R. Vio}

   \institute{Chip Computers Consulting s.r.l., Viale Don L.~Sturzo 82,
              S.Liberale di Marcon, 30020 Venice, Italy\\
              ESA-VILSPA, Apartado 50727, 28080 Madrid, Spain\\
              \email{robertovio@tin.it}
         \and
              Department of Mathematical and Computer Sciences, Colorado School of Mines,
                  Golden CO 80401, USA \\
              \email{ltenorio@Mines.EDU}
         \and
             ESA-VILSPA, Apartado 50727, 28080 Madrid, Spain\\
             \email{willem.wamsteker@esa.int}
             }

\date{Received .......................; accepted ..........................}

\abstract{Point-source contamination in high-precision Cosmic Microwave Background (CMB) maps severely affects
the precision of cosmological parameter estimates. Among the methods that have been proposed for
source detection, wavelet techniques based on ``optimal'' filters have been proposed
\citep[e.g.,][]{san01}. In
this paper we show that these filters are in fact only restrictive cases of a more general class of matched
filters that optimize signal-to-noise ratio and that have, in general, better source detection
capabilities, especially for lower amplitude sources.
These conclusions are confirmed by some numerical experiments.
\keywords{Methods: data analysis -- Methods: statistical}
}
\titlerunning{Optimal Detection of Sources}
\authorrunning{R. Vio, L. Tenorio \& W. Wamsteker}

\maketitle

\section{Introduction}

The separation of different physical components is an important issue in the analysis
of Cosmic Microwave Background (CMB) data. Among the foreground components,
point-sources deserve especial attention
given their extreme non-Gaussianity and highly variable spectral index. A brief summary
of different methods can be found in \citet{san01} (henceforth \shm).

Methods used to detect point sources should act as high-pass filters
to detect the high frequency structure introduced by the sources
in the CMB data while masking other foreground contamination (like dust, synchrotron and free-free emission)
characterized by lower frequencies. In principle, the good space/frequency characteristics of wavelets
should make these functions an attractive
choice for such tasks
\citep[e.g.,][ and references therein]{cay00,viea,vieb}. Indeed, wavelets have been proved optimal for detection
of point like singularities -- at least for one-dimensional signals.
Point sources in CMB maps, however, are not point singularities because the signal is smoothed by the beam of the
antenna; sources are expected to have the shape of the antenna's pattern. The question is
then how to include this beam profile information in the wavelet analysis. {\shm} considered this question and,
were lead to define optimal pseudo-filters for
source detection in CMB maps. However, we show that nothing seems to be gained with these wavelet filters and that other
simpler techniques lead to better source detection methods.

We first set up the basic framework. Although the detection of point-like sources in CMB maps is a
two-dimensional problem, we present our arguments in $R^n$, as in \shm, because
the same methods may be used in other applications.

The sources are assumed to be point-like signals convolved with the beam of the measuring instrument and are
thus assumed to have a known profile $\tau(\xb)$. The signal $y(\xb)$, $\xb\in R^n$, is modeled as
\begin{equation}\label{datamodel}
y(\xb) = \sum_{j} s_j(\xb) + z(\xb)
\end{equation}
where
\begin{equation}
s_j(\xb)=A_j\, \tau(\xb -\xb_j),
\end{equation}
$A_j$ and $\xb_j$ are, respectively, unknown source amplitudes and locations, and $z(\xb)$ is a zero-mean background
with power-spectrum $P(\qb)$
\begin{equation} \label{eq:powerz}
{\rm E}\,[\,z(\qb)\, z^*(\qb')\,] = P(\qb) ~\delta^n(\qb - \qb').
\end{equation}
Henceforth ${\rm E}[\cdot]$ and ``$~{}^*~$'' will denote the expectation and complex conjugate operators, respectively,
$\delta^n(\qb - \qb')$ the $n$-dimensional Dirac distribution, and $z(\qb)$ the Fourier transform of $z(\xb)$
\begin{equation}
z(\qb) = \int_{-\infty}^{+\infty} z(\xb) ~{\rm e}^{- i \qb \cdot \xb} ~\db\xb.
\end{equation}
To properly remove the point sources from the signal we need to estimate the locations  $\{\xb_j\}$ and
amplitudes (fluxes) $\{A_j\}$
of the sources.

A classical method used to estimate source locations is based on identifying peaks in the cross-correlation
function
\begin{equation} \label{eq:correlation}
c(\xb) = \int_{-\infty}^{+\infty} y(\xb + \bb) ~\tau(\bb) ~\db \bb.
\end{equation}
The rationale is that $c(\xbo)$ measures the similarity between the source profile with a section of
$y(\xb)$ centered at $\xbo$; a peak in $c(\xbo)$ is an indication of a source signal at $\xbo$.
Eq. (\ref{eq:correlation}) is a filtering of the signal $y(\xb)$ with a
filter $\tau(\bb)$ that amplifies the characteristic frequencies of the source.
Once the sources have been located, their amplitudes can be estimated by means of classical
fitting procedures like least squares.

The cross-correlation function (\ref{eq:correlation}) does not take
into account the background characteristics.
This is a great disadvantage in cases where the power spectrum $P(\qb)$ is known or a good estimate is available.
In Sects. \ref{sec:design} and \ref{sec:pseudo} we consider other methods that take into account this information
and that may be considered extensions of the cross-correlation method.

The basic procedure we consider is as follows. The signal is first filtered to enhance the sources
with respect to the background. This is done by cross-correlating the signal $y(\xb)$ with a filter $\Phi$ as in
(\ref{eq:correlation}) (with $\Phi$ in place of $\tau$).
The source locations are then determined by selecting the peaks in the filtered signal
that are above a selected threshold. Finally, the source amplitudes are estimated with the
values of the filtered signal at the estimated locations. The question we consider first is the selection of an
optimal filter $\Phi$ for such procedure.

\section{Designing an optimal filter}
\label{sec:design}
The optimality criteria we use are based on the following assumptions (for futher justification of
these assumptions see \shm). The source profile and background spectrum
are known. The profile is spherically symmetric, characterized by a scale $R_s$, and the background is assumed to
be isotropic.
In this case, we can write
$s(\xb) \equiv s(x)$, where $x = \| \xb \|$, and $P(\qb) \equiv P(q)$ for $q = \| \qb \|$. In addition,
source overlap is
assumed negligible.

We consider the general family of spherically symmetric filters $\Phi(\xb; \bb)$ of the form
\begin{equation}\label{eq:Phifam}
\Phi(\xb;\bb) = \phi(\, \| \xb - \bb \|\, )
\end{equation}
with Fourier transform $\phi(q)$. The filtered field is
\begin{equation} \label{eq:phiweq}
\begin{split}
w(\bb;\phi)& =
\int_{-\infty}^{+\infty} y(\xb) ~\Phi(\xb; \bb) ~\db\xb \\
&= \int_{-\infty}^{+\infty} y(\qb) ~ \phi(q) ~{\rm e}^{i \qb \cdot \bb} ~\db\qb,
\end{split}
\end{equation}
where $ y(\qb)$ and $ \phi(q)$ are, respectively, the Fourier transforms of $y(\xb)$ and $\phi(\xb)$.
The mean and variance of $w(\bb;\phi)$ are
\begin{equation}
\begin{split}
\mu(\bb;\phi) &= {\rm E}\,w(\bb;\phi) \\ &= \alpha \int_0^{+\infty}\!\! q^{n-1} ~s(q)\, \phi(q)\,
{\rm e}^{i \qb \cdot \bb} dq;
\end{split}
\end{equation}
\begin{equation} \label{eq:sigma}
\begin{split}
\sigma^2(\phi) &= {\rm E}\,w^2(\bb,\phi)  - \mu(\bb;\phi)^2 \\
& =\alpha \int_{0}^{+\infty} \!\! q^{n-1}~P(q)\, \phi(q) ~dq,
\end{split}
\end{equation}
where $\alpha= 2\, \pi^{n/2}\,\Gamma^{-1}(n/2)$.

The first constraint on the filter concerns the second stage of the procedure; the source locations are assumed known
and the objective is to estimate the amplitudes. Given the assumed distance between sources, it is enough to consider
a field $y(\xb)$ as in (\ref{datamodel}) with a single source at the origin, $s(\xb) = A\,\tau(\xb)$.
To estimate its amplitude we ask that $w(\Ob;\phi)$ be an unbiased estimator of $A$ -- i.e.,
$\mu(\Ob;\phi) = A$ -- so that $\phi$ is required to satisfy the equation
\begin{equation} \label{eq:constr1}
\int_0^{+\infty}\!\! q^{n-1} \tau(q)\, \phi(q) ~dq = \frac{1}{\alpha}.
\end{equation}
To enhance the magnitude of the source relative to the background we determine the filter $\Phi$ that minimizes
the variance $\sigma^2(\phi)$. This has the effect of maximizing, among unbiased estimators,
the detection level
\begin{equation} \label{eq:detection}
\mathcal{D}(\phi)=\frac{\mu(\Ob;\phi )}{\sigma(\phi)},
\end{equation}
which measures the capability of the filter to correctly detect a source at the prescribed location (see \shm).

Since $\Phi$ is chosen so that $w(\Ob;\phi)$ is a minimum variance linear -- in $y(\xb)$ -- unbiased
estimator of $A$, it follows that (Gauss-Markov theorem) $w(\Ob;\phi)$ is the (generalized) least squares estimate
of $A$ achieved by the filter
\begin{equation} \label{eq:filter2}
\phi(q) = \frac{1}{\alpha a} ~ \frac{\tau(q)}{P(q)},\qquad a \equiv \int_0^{+\infty}q^{n-1} \frac{\tau^2}{P} ~dq,
\end{equation}
with minimum variance
\begin{equation} \label{eq:sigma2}
\sigma^2(\phi) = \frac{1}{\alpha a}.
\end{equation}
Filter (\ref{eq:filter2}) is a particular case of a well known class of filters, known as
{\it matched filters} in the engineering literature, that are designed to optimize signal-to-noise ratio
\citep[e.g.,][]{kozma,pratt91}. The arguments in this section also show that the filter introduced by
\cite{teg98} is a particular case of the matched filter (\ref{eq:filter2}) with $P(q)$ representing
the background power
spectrum before the smoothing of the antenna.

For white noise, $P(q) = {\rm const} = D$, filter (\ref{eq:filter2}) simplifies to
\begin{equation}\label{phiwn}
\phi(q) = \frac{1}{\alpha a D} ~ \tau(q)
\end{equation}
which, up to a constant factor, is identical to the filter used in the classical cross-correlation function
(\ref{eq:correlation}). This provides a justification for the use of the cross-correlation for
source detection in white noise.

Having recognized $w(\Ob;\phi)$ as a least squares estimator of $A$, we close this section with some remarks from
least squares methodology that we consider relevant. First note that, regardless of the spectrum $P(q)$, the source
profile $\tau(q)$ properly normalized, that is
$\phi(q) = \tau(q)/K$ for $K=\alpha\,\int q^{n-1} \tau(q)\,\phi(q)\, {\rm e}^{i\qb\cdot\bb}\,dq$, also provides an
unbiased estimator $w(\Ob;\phi)$ of $A$. This is the (ordinary) least squares estimate that does not take into
account the covariance of the background; it is unbiased but not minimum variance. However, it is well known
that when the covariance is actually estimated from the data, the ordinary least squares estimate may
be better than the generalized one \citep[e.g.,][]{dra98}. In other words,
uncertainities in the spectrum estimates may lead to worse amplitude estimates than those obtained with
the simpler cross-correlation filter. Uncertainties in the spectrum will also affect the selection of a
detection threshold.

Note also that the unbiasesness of $w(\Ob;\phi)$ as an estimator of $A$ depends on knowing the correct source
location, it is not necessarily unbiased once the source locations are estimated from the data. This is shown
in Sect. \ref{sec:numexp}.

\section{Pseudo-filters}
\label{sec:pseudo}
In the pseudo-filter approach of {\shm} the filters are of the form
(\ref{eq:Phifam}) with an additional scale
dependence
\begin{equation} \label{eq:psi}
\Psi(\xb; R, \bb) = \frac{1}{R^N}\, \psi \left( \frac{\| \xb - \bb \|}{R} \right),
\end{equation}
for some spherically symmetric function $\psi$. From (\ref{eq:psi}) we get a set of wavelets of different
scales $R$ and shifts $\bb$ provided the regularity conditions $\tau(q)/P(q) \rightarrow 0$,
$(1/P)(d\tau /d \ln q) \rightarrow 0$ for $q \rightarrow 0$ are satisfied.
The filtered field $w(\bb,R;\psi)$ at scale $R$ is defined as in (\ref{eq:phiweq})
but with $\psi(Rq)$ in place of $\phi(q)$.

To determine an optimal filter $\psi$, \shm\  minimize the variance of the filtered field subject
to two constraints: First, $w(\Ob,R_0;\psi)$ is required to be, as in the previous section,
an unbiased estimator of $A$ for some known $R_0 \approx R_s$. For the second constraint $\psi$ is selected so
that $\mu(\Ob,R;\psi)$
has a local maximum at scale $R_0$. This constraint translates to
\begin{equation} \label{eq:constr2}
\int_0^{+\infty} q^{n-1} \tau(q) \,\psi(R_0 q) \left(n + \frac{d \ln \tau}{d \ln q} \right)~dq = 0.
\end{equation}
Minimizing $\sigma^2(R_0;\psi)$ with the two constraints yields the filter (\shm)
\begin{equation}
\psi(R_0 q) = \frac{1}{\alpha ~\Delta} ~\frac{\tau(q)}{P(q)} \left[ nb + c -(na+b)\,
\frac{d \ln \tau(q)}{d \ln q} \right],
\end{equation}
where $\Delta = ac - b^2$,
\begin{equation}
\begin{aligned}
b &\equiv \int_0^{+\infty} q^{n-1} \frac{\tau}{P} ~\frac{d\tau}{d\ln q}, \\
c &\equiv \int_0^{+\infty} q^{n-1} \frac{1}{P} \left( \frac{d \tau}{d \ln q} \right)^2 ~dq,
\end{aligned}
\end{equation}
and $a$ is as in (\ref{eq:filter2}). This filter provides a field of variance
\begin{equation} \label{eq:sigma1}
\sigma^2(R_0;\psi) = \frac{n^2 a + 2 n b + c}{\alpha \Delta}.
\end{equation}

The estimator of the amplitude $A$ obtained with this filter is again linear and unbiased and therefore,
by the optimality of least squares, $\sigma^2(R_0;\psi)\geq \sigma^2(\phi)$ regardless of the source
profile and background spectrum. This means that
the detection level of $\Phi$ is at least as high, or higher, than that achieved with $\Psi$.

\subsection{Is $\Psi$ optimal for source detection?}

We have seen that the second constraint (\ref{eq:constr2}) does not increase the detection level when the source
locations
are known. But this is not surprising since the constraint is defined to take advantage of the known source
scale to help determine source locations. We will show that even when source
location uncertainty is taken into account, (\ref{eq:constr2}) does not improve on
the simpler filter $\Phi$ based on the single contraint (\ref{eq:constr1}). In other words, enough information
about the scale of the source is already included in the derivation of the matched filter.
Therefore, it seems that nothing is gained with wavelet methods.

In principle, as explained by \shm, constraint
(\ref{eq:constr2}) can be used to test if a detection
corresponds to a true source by checking for its maximum at scale $R_0$. But finding a spike at the correct
scale is not enough to make sure it is not a spurious noise artifact, we have to follow it across scales to
make sure that it scales appropriately. In any case, a similar test can also be designed for the matched filter:
given that the source
profile is assumed known, the matched filter leads to an amplitude versus scale dependence (see example in Sect.
\ref{sec:gex}) that can be
determined and used for such a test. This functional dependence provides more
information than the existence of a local maxima and should, in the presence of noise, lead to better
location estimates.

\section{Examples}
To compare the theoretical performances of the filters $\Phi$ and $\Psi$ we use the {\it gain}
as defined by \shm
\begin{equation}
\begin{split}
g(\psi,\phi) &= {\Dc}(\phi)/\Dc(\psi,R_0) \\ &=\sigma(R_0;\psi) / \sigma(\phi),
\end{split}
\end{equation}
where $\Dc(\psi,R_0)$ is as in (\ref{eq:detection}) but defined for $\Psi(\xb;R_0,\Ob)$.
We start with a specific example with Gaussian profiles and power-law spectra.

\subsection{Gaussian sources and $P(q) = D q^{-\gamma}$}
\label{sec:gex}

Gaussian sources
\begin{equation}
\tau(q)=\theta^n\, {\rm e}^{-(q \theta)^2/\,2},
\end{equation}
where $\theta$ is the ``standard deviation'' defining the scale,
provide an important family of source profiles.
Indeed, in many practical
applications the instrument's profile, and thus the point sources, are effectively characterized by Gaussian profiles.
For the noise process we take the power-law
spectrum $P(q) = D q^{-\gamma}$. This family of spectra can be used to locally approximate spectra of other homogeneous
processes and has as special cases white ($\gamma=0$) and
$1/f$ ($\gamma=1$) noise processes.

\begin{figure}
        \resizebox{\hsize}{!}{\includegraphics{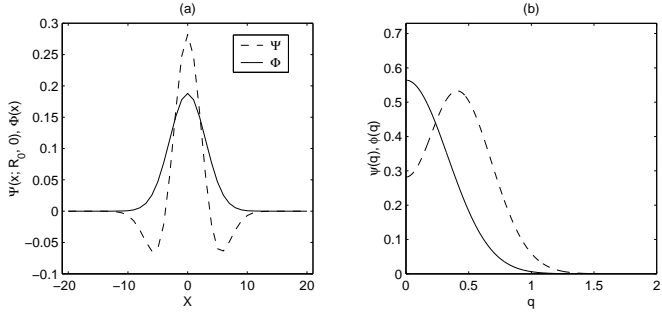}}
        \caption{ Filters $\Psi(x; R_0, 0)$ and $\Phi(x)$ for $n=1$ and white noise.
        Panel {\bf (a)} shows the filters in spatial domain and panel {\bf (b)} shows their power-        spectra.}
        \label{fig:filt_w}
\end{figure}
\begin{figure}
        \resizebox{\hsize}{!}{\includegraphics{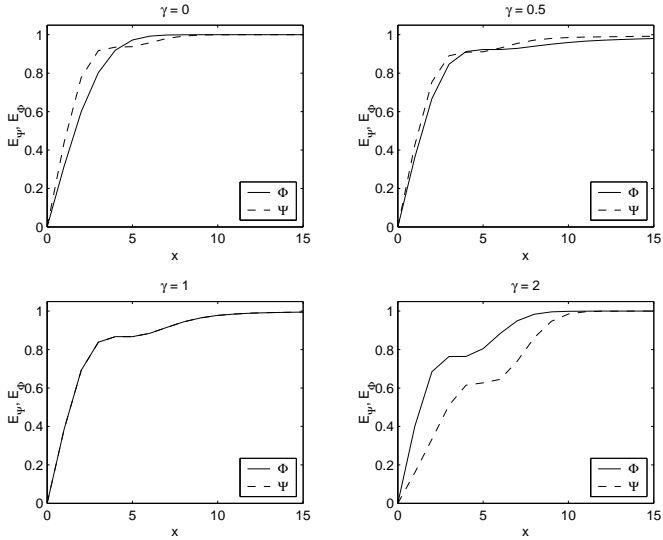}}
        \caption{ Cumulative energies $E_{\psi}$ and $E_{\phi}$ of
        $\Psi(x; R_0, 0)$ and $\Phi(x)$, respectively, for $n=1$ and different values of $\gamma$       (see text).}
        \label{fig:filt_energy1}
\end{figure}
\begin{figure}
        \resizebox{\hsize}{!}{\includegraphics{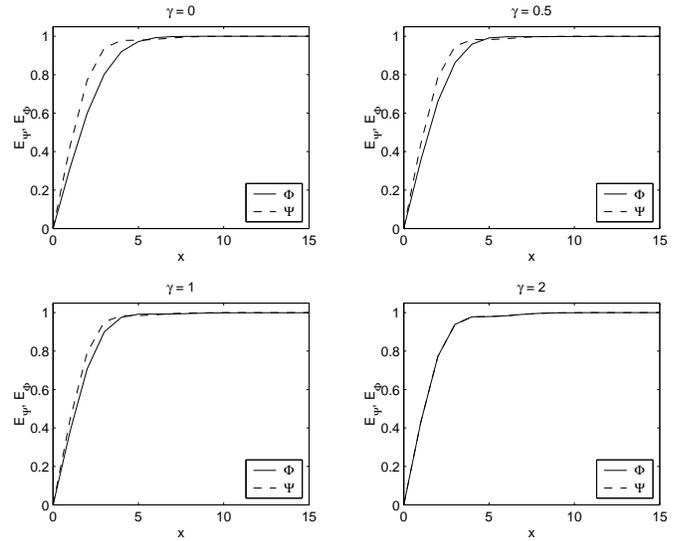}}
        \caption{ Same as Fig. \ref{fig:filt_energy1} but with $n=2$.}
        \label{fig:filt_energy2}
\end{figure}

\begin{table*}[t]
\begin{center}
\begin{tabular}{cccccccccccc}
\hline
\hline
\\
\multicolumn{1}{c}{} & \multicolumn{3}{c}{$S/N=1$} &  \multicolumn{1}{c}{} & \multicolumn{3}{c}{$S/N=2$} &
\multicolumn{1}{c}{} & \multicolumn{3}{c}{$S/N=3$} \\
\cline{2-4} \cline{6-8} \cline{10-12}   \\
${\rm Filter}$ & $N_c$ & $N_i$ & $A$ & & $N_c$ & $N_i$ & $A$ & & $N_c$ & $N_i$ & $A$ \\
\hline
\\
$\Psi$ &23 & 10 & 1.83 & & 87& 10 & 1.10 & & 100& 10 & 0.97\\
          &(15) &  (5) & (1.94) & & (80) & (5) & (1.14) & & (100)& (5) & (0.98) \\
$\Phi$ &31 & 5  & 1.55 & & 97& 5  & 1.02 & & 100& 5 &  0.98\\
\\
\hline
\\
\end{tabular}
\caption{Results of the numerical simulations concerning the detection capabilities of the matched and the
pseudo-filters (see text) in the one dimensional case and for a white noise background. $N_c$, $N_i$ are,
respectively, the average number of correct and incorrect detections.
Five hundred simulations were carried out using $100$ gaussian sources of amplitude $A=1$ and scale $\theta=3$,
regularly distributed along an array of about $16000$ elements. The $3\sigma$ detection level has been
determined by filtering, via $\Phi$ and $\Psi$, independent realizations of the background  process. The
amplitude estimates are based on the average amplitudes of all detected sources. The S/N is defined by
$A/\sigma_e$ where $\sigma_e$ is the standard deviation of the noise process.
The second row for $\Psi$ corresponds to results with a threshold ($>3\sigma$) chosen to achieve the same
average number of incorrect detections as $\Phi$.\label{tbl:white1}}
\end{center}
\end{table*}

\begin{table*}[t]
\begin{center}
\begin{tabular}{cccccccccccc}
\hline
\hline
\\
\multicolumn{1}{c}{} & \multicolumn{2}{c}{$S/N=1$} &  \multicolumn{1}{c}{} & \multicolumn{2}{c}{$S/N=2$} &
\multicolumn{1}{c}{} & \multicolumn{2}{c}{$S/N=3$} \\
\cline{2-3} \cline{5-6} \cline{8-9}   \\
${\rm Filter}$ & $P_c$ & $A$ & & $P_c$ & $A$ & & $P_c$   & $A$ \\
\hline
\\
$\Psi$ & 0.89 & 1.14 & & 1.00 & 1.01 & & 1.00 & 1.00\\
$\Phi$ & 0.99 & 1.03 & & 1.00 & 1.00 & & 1.00 & 1.00\\
\\
\hline
\\
\end{tabular}
\caption{Results of two-dimensional simulations with a $128\times 128$ grid and white noise.
$A$ is as in Table \ref{tbl:white1}.
$P_c$ is the estimated probability of correctly detecting a source
keeping the rate of incorrect detection at the same level as in Table \ref{tbl:white1}.
Here $\Psi$ coincides with the Mexican hat wavelet.
\label{tbl:white2}}
\end{center}
\end{table*}

\begin{table*}[t]
\begin{center}
\begin{tabular}{cccccccccccc}
\hline
\hline
\\
\multicolumn{1}{c}{} & \multicolumn{2}{c}{$S/N=1$} &  \multicolumn{1}{c}{} & \multicolumn{2}{c}{$S/N=2$} &
\multicolumn{1}{c}{} & \multicolumn{2}{c}{$S/N=3$} \\
\cline{2-3} \cline{5-6} \cline{8-9}   \\
${\rm Filter}$ & $P_c$ & $A$ & & $P_c$ & $A$ & & $P_c$   & $A$ \\
\hline
\\
${\rm MH}$   & 0.38 & 1.79 & & 0.93 & 1.11 & & 1.00 & 1.03\\
$\Psi$ & 0.40 & 1.78 & & 0.93 & 1.11 & & 1.00 & 1.03\\
$\Phi$ & 0.47 & 1.60 & & 0.97 & 1.07 & & 1.00 & 1.02\\
\\
\hline
\\
\end{tabular}
\caption{Results of two-dimensional simulations with a $128\times 128$ grid and $1/f$ noise.
$A$ is as in Table \ref{tbl:white1}.
$P_c$ is the estimated probability of correctly detecting a source
keeping the rate of incorrect detection at the same level as in Table \ref{tbl:white1}.
${\rm MH}$ stands for Mexican hat wavelet.
\label{tbl:1f}}
\end{center}
\end{table*}

For a Gaussian profile and a power-law spectrum, Eq. (\ref{eq:filter2}) leads to
\begin{equation}\label{eq:phi}
\phi(q) = \frac{\Gamma(n/2)}{\Gamma(m)} ~ \frac{(q \theta)^{\gamma}}{\pi^{n/2}}
~{\rm e}^{- (q \theta)^2/\,2},
\end{equation}
where $m= (n+\gamma)/2$.
>From (\ref{eq:sigma2}) and (\ref{eq:sigma1}) we obtain the following variance and gain
\begin{equation}\label{eq:varg}
\begin{aligned}
\sigma^2(\phi) & = \frac{D}{\theta^{n-\gamma} ~\pi^{n/2}} ~\frac{\Gamma(n/2)}{\Gamma(m)}, \\
\qquad g(\psi,\phi) & = \left[\, 1 + \frac{(n - \gamma)^2}{4 m} \,\right]^{1/2}.
\end{aligned}
\end{equation}
We see that $g \geq 1 $, which shows that $\phi$ has a higher detection level than $\psi$.
For $n=\gamma$ ---for example,
one-dimensional process with $1/f$ noise--- the two filters lead to the same detection
levels. This is expected since $\phi(q) = \psi(Rq)$ for $n=\gamma$ (compare Eq. (\ref{eq:phi}) with Eq. (24)
in {\it SHM}).
In other words, the second constraint is just
redundant in this case. Note also that $\phi(q)$ is the Mexican hat wavelet for $n=2$ and $
\gamma=2$. This results show that, contrary to what has been claimed before \citep[e.g.,][]{cay00},
the optimality of the Mexican hat wavelet does depend on the background spectrum.

\subsection{A numerical experiment}
\label{sec:numexp}
We have presented theoretical arguments showing that $\phi(q)$ has a better source detection
capability than $\psi(Rq)$ when the source location is known. We now confirm that $\phi$ is still better
when the uncertainity of source location estimates is taken into account. To compare with the results in \shm, it
is enough to consider a simple example with one-dimensional Gaussian sources and white noise ($\gamma=0$)
(we already know that $\phi = \psi$ for $n=1$ and $1/f$ noise).
In this case (\ref{phiwn}) and (\ref{eq:varg}) become, respectively,
\begin{equation}\label{eq:phiexp}
\begin{aligned}
\phi(q) & = \frac{1}{\sqrt{\pi}} ~ {\rm e}^{-(q \theta)^2/\,2}, \\
g(\psi,\phi) & = \left( \frac{3}{2} \right)^{1/2} > 1.
\end{aligned}
\end{equation}
That is, the detection level of $\phi$ is about 20\% larger than that of $\psi$.
The amplitude dependence on $\theta$ of a Gaussian source of scale $R_s$ filtered with $\phi$ is
\begin{equation}
A(\theta) = \frac{\sqrt{2}\,R_s}{(\,R_s^2 +\theta^2\,)^{1/2}}\,\,A.
\end{equation}
A fit to this dependence can be used to determine if a detection corresponds to a source of the correct scale,
just as large wavelet coefficients would be tracked across different wavelet scales.

Fig. \ref{fig:filt_w} shows the filters $\Phi(\xb)$ and $\Psi(\xb; R_0, \Ob)$ and their corresponding
Fourier transforms. It shows that the two filters are quite different. For example,
to provide filtered sources with the scale $R_0$, $\psi$ has to pass higher
frequencies than $\phi$. This can be a problem for signals contaminated by high frequency noise.

Table \ref{tbl:white1} shows the average number of correct and incorrect detections obtained with
$\Psi$ and $\Phi$ and a
fixed $3\sigma$ threshold for signal-to-noise (S/N) ratios equal to 1, 2 and 3. The two filters give equivalent
results for higher S/N sources. We see that $\phi$ leads to a higher number of
correct detections and a lower number of
incorrect ones. But a proper comparison should take into account that the filters require different
thresholds. The second row for $\Psi$ shows the corresponding results when the threshold is chosen
to lead to the same average number of rejections as $\Phi$. For low S/N sources $\Phi$ leads again to a
higher number
of correct detections while for larger S/N they give similar results.

To compare amplitude estimates that include location uncertainty, we take the average of the amplitudes (since all
the generated sources have the same amplitude) of all detections. The results are shown in Table \ref{tbl:white1}.
The errors in the amplitudes are of the order of 0.5\% or less. We see that amplitude estimates are biased when the
source locations are estimated, and that the bias is larger for $\Psi$. For low S/N the amplitude is
overestimated
because high peaks are easier to detect and noise peaks are incorrectly classified as sources. For high S/N sources
we also have centering problems but this time the smaller amplitude in the noise peaks leads to underestimated
source amplitudes. We can draw similar conclusions from the results of two-dimensional simulations
shown in Tables \ref{tbl:white2} and \ref{tbl:1f};

To conclude, note that the spatial support of the filters is an important factor when the assumption of
nonoverlapping filters is
invalid.
The support must be small compared to the distance between the sources.
Figs. \ref{fig:filt_energy1}-\ref{fig:filt_energy2} show (for $n=1,2$) the cumulative energies
\begin{equation}
{\rm E}_{\psi}(x) = \frac{\int_{0}^{x} \Psi^2(b; R_0, 0) db}{\int_{0}^{\infty} \Psi^2(b; R_0, 0) db}
\end{equation}
and
\begin{equation}
{\rm E}_{\Phi}(x) = \frac{\int_{0}^{x} \Phi^2(b) db}{\int_{0}^{\infty} \Phi^2(b) db},
\end{equation}
as a function of $\gamma$ for a background spectrum $P(q) = D q^{-\gamma}$.
These functions measure the ``energy concentration'' of the filters and
provide information about their ``spatial'' support. The figures show that also in this respect
$\Phi(x)$ has similar, if not better, characteristics than $\Psi(x; R_0, 0)$. In particular, the filter $\Phi$ has a
tighter support for faster decaying noise spectra.

\subsubsection{Selecting the detection level}
We briefly justify our selection of $3\sigma$ level.
This threshold should be chosen large enough to reduce the number of false detections but small enough not to
miss too many sources. To properly choose a threshold we have to understand the statistics of
local maxima above a chosen level. For a general homogeneous (Gaussian) random field
this is a difficult question \citep[some asymptotic results can be found in][]{adler} but simulations can be
carried out when the background power spectrum is known. Our simulations showed that
the traditional $5\sigma$ level is too conservative for the signal lengths used in the examples.
If $P_M$ is the proportion of local maxima above $3\sigma$ for a field without sources. We found that the
probability
that $P_M$ is higher than 0.001 is about $9\%$ for the signal filtered with $\Psi$ and less than $10^{-5}$ for
the signal filtered with $\Phi$.

\section{Summary and Conclusions}

We have revisited the problem of estimating point sources of known profile in an isotropic background.
The methods we considered are based on two basic interrelated procedures: source detection
by thresholding of local maxima,
and amplitude estimation by linear filtering. We have compared the effects of different constraints on the selection of
an optimal filter.

The first constraint is typical in matched filter methodology where S/N
is maximized. The optimal filter provides unbiased least squares estimates of
source amplitudes at known source locations. By the optimality of least squares, these amplitude
estimates can not be improved with any other unbiased linear filter.
However, uncertainities in source locations
introduce a bias in amplitude estimates. Amplitudes are overestimated at low S/N and underestimated at
high S/N.
A second constraint introduced by {\shm} is designed to improve estimates of source locations by maximizing
amplitudes
at the correct scale. We found that this constraint does not lead to better estimates as compared
to those obtained with a simpler matched filter, especially for lower S/N sources. For high S/N sources,
as it was observed by \cite{cay00},
the results of the two methods are the same.
These results contradict previous optimality studies of wavelet-based filters for detection of point-sources
of known profile in an isotropic CMB background of known spectrum.

\clearpage


\begin{thebibliography}{}
\bibitem[Adler (1981)]{adler} Adler, R. J. 1981, The Geometry of Random Fields (Wiley, New York)
\bibitem[Cay\'{o}n et al. (2000)]{cay00} Cay\'{o}n, L. et al. 2000, MNRAS, 315, 757
\bibitem[Draper \& Smith (1998)]{dra98} Draper, N. R. \& Smith, H. 1998, Applied Regression Analysis (Wiley, New York)
\bibitem[Kozma \& Kelley (1965)]{kozma} Kozma, A. \& Kelly, D. L. 1965, Appl. Opt., 4, 387
\bibitem[Pratt (1991)]{pratt91} Pratt, W. K. 1991, Digital Image Processing (Wiley, New York)
\bibitem[Sanz et al. (2001)]{san01} Sanz, J. L., Herranz, D., \& Martinez-Gonzalez, E. 2001,
ApJ, 552, 484 ({\it SHM})
\bibitem[Tegmark \& Oliveira-Costa (1998)]{teg98} Tegmark, M. \& Oliveira-Costa, A. 1998, ApJ, 500L, 83
\bibitem[Vielva et al. (2001a)]{viea} Vielva, P. et al. 2001, MNRAS, 326, 181
\bibitem[Vielva et al. (2001b)]{vieb} Vielva, P. et al. 2001, MNRAS, 328, 1
\end{thebibliography}
\end{document}